\documentclass[sn-mathphys]{sn-jnl}
\usepackage{amsmath,amssymb,amsfonts}
\usepackage{multirow}
\usepackage{multicol}
\usepackage{subfigure}
\usepackage{subcaption}
\usepackage{caption}
\usepackage{graphicx}
\usepackage{tabularx}

\jyear{2023}%

\theoremstyle{thmstyleone}%
\theoremstyle{thmstyletwo}%

\theoremstyle{thmstylethree}%

\begin{document}

\title[Siamese Architecture for LOB]{An Efficient deep learning model to Predict Stock Price Movement Based on Limit Order Book}


\author*{\fnm{Jiahao} \sur{Yang}}\email{yangjiahaostu@gmail.com}
\author{\fnm{Ran} \sur{Fang}}\email{fangran@hccl.ioa.ac.cn}
\author{\fnm{Ming} \sur{Zhang}}\email{zhangming.peter@gmail.com}
\author{\fnm{Jun} \sur{Zhou}}\email{zhoujun@hccl.ioa.ac.cn}

\affil{\orgdiv{The Institute of Acoustics of the Chinese Academy of Sciences}, \orgaddress{ \city{Beijing}, \postcode{100190},  \country{China}}}

\affil{\orgdiv{University of Chinese Academy of Sciences}, \orgaddress{ \city{Beijing},  \country{China}}}



\abstract{In high-frequency trading (HFT), leveraging limit order books (LOB) to model stock price movements is crucial for achieving profitable outcomes. \textcolor{black}{However, this task is challenging due to the high-dimensional and volatile nature of the original data.} Even recent deep learning models often struggle to capture price movement patterns effectively, particularly without well-designed features. We \textcolor{black}{observed that raw LOB data exhibits inherent symmetry between the ask and bid sides,} and the bid-ask differences demonstrate greater stability and lower complexity \textcolor{black}{compared to the original data. Building on this insight}, we propose a novel approach in which \textcolor{black}{leverages} the Siamese architecture to enhance the performance of existing deep learning models. The core idea involves processing the ask and bid sides separately using the same module \textcolor{black}{with shared parameters}. We \textcolor{black}{applied} our Siamese-based methods to several widely used strong baselines and validated \textcolor{black}{their effectiveness using data from} 14 military industry stocks in the Chinese A-share market. Furthermore, we integrated multi-head attention (MHA) mechanisms with the \textcolor{black}{Long Short-Term Memory (LSTM)} module to investigate its role in modeling stock price movements. Our experiments used raw data and widely used Order Flow Imbalance (OFI) features as input with some strong baseline models. The results show that our method improves the performance of strong baselines in over 75$\%$ of cases, excluding the \textcolor{black}{Multi-Layer Perception (MLP)} baseline, which performed poorly and is not considered practical. Furthermore, we found that \textcolor{black}{Multi-Head Attention} can enhance model performance, \textcolor{black}{particularly over shorter forecasting horizons.}}





\maketitle
\section{Introduction}

The rise of algorithmic trading has \textcolor{black}{made} High-Frequency Trading (HFT) one of the most widely adopted trading techniques in the stock market, gaining increasing popularity over time. By 2016, the proportion of HFT in the U.S. and European equity markets had reached roughly 55\% and 40\% of trading volume, respectively \citep{1}. The Limit Order Book (LOB) is a dynamic data structure that serves as a comprehensive record of all the orders submitted to a financial market. It is a valuable source of information for market participants, as it is often used to predict future price movements and make trading decisions.

Previous approaches have predominantly relied on statistical methods, where researchers often propose hypotheses to model the trading process, such as regarding the arrival of all types of orders as a Poisson process, estimating model parameters with transaction data \citep{5, 7} and \textcolor{black}{and traders dynamically choosing between limit and market orders \citep{6}.} Additionally, some work estimates the limit order book parameters in Kalman filter methods to reveal the pattern in the limit order book \citep{8}. Nonetheless, the intricate and stochastic nature of stock markets imposes significant limitations on the performance of these methods. 

In response to these challenges, recent research has increasingly explored deep learning-based approaches for stock price forecasting, employing strategies such as curriculum learning to improve training efficiency and generalization \citep{curriculm0,curr1,curriculum}, modeling data heterogeneity and distribution shifts \citep{hete2,hete3,heteroscedastic}, designing novel ranking-based loss functions \citep{knowledge1,knowledge2,knowledge3,knowledge}, and developing hybrid architectures to enhance predictive performance \citep{zhao2022forecasting,zhao2024stock,CNN-GRU}.

Building on these advancements, machine learning methods have also gained prominence specifically in modeling Limit Order Book (LOB) data, offering the potential to capture its complex structure and dynamics more effectively. Support Vector Machine (SVM) is one of the most popular machine learning methods to predict \textcolor{black}{price movements} based on LOB data, either as a classifier to predict the trend of price changes \citep{9} or as a regressor to predict the specific value of price changes \citep{10}. Due to their robustness in noisy environments, bagging-based methods such as Random Forest also work well in this task \citep{11}. Meanwhile, some gradient-based boosting algorithms perform well at capturing hidden trading patterns, such as XGBoost \citep{12} and CatBoost \citep{13}. Some unsupervised algorithms, such as least-mean-squares (LMS) and linear discriminant analysis (LDA) \textcolor{black}{are useful for guiding feature selection from handcrafted features \citep{14}.} In addition, a variety of hand-designed features are necessary. For instance, time-insensitive features describe the trading status of each depth level, while time-sensitive features capture the average intensity of trades \citep{15}.

The main obstacles encountered are the dynamic and high-dimensional nature of LOB data, so deep learning models have attracted more and more attention in recent years due to their ability to learn representations automatically. Some researchers \textcolor{black}{applied} deep learning models \textcolor{black}{to} a U.S. High-Frequency Trading (HFT) database comprising billions of market quotes and transactions. Their findings demonstrate that the deep learning model \textcolor{black}{outperforms} asset-specific linear and nonlinear models \citep{16}. This suggests that the deep learning model is capable of capturing universal and \textcolor{black}{stable} trading patterns. 
Researchers have gradually proposed many \textcolor{black}{innovative} network architectures to process the LOB data and predict stock prices, including Multi-Layer Perceptron (MLP) \citep{16}, Convolutional Neural Network (CNN) \citep{17,18}, Long Short-Term Memory (LSTM) \citep{19}, Gated Recurrent Unit (GRU) \citep{20}, CNN-LSTM \citep{22}. Furthermore, deep learning models are used to extract features from the LOB data to \textcolor{black}{mitigate} the impact of noise. Leangarun et al. proposed utilizing the unsupervised autoencoder (AE) and generative adversarial networks to process the trading data \citep{23}. \textcolor{black}{Huang et al. combined deep neural networks with bag-of-features (BoF) models to predict price movements \citep{24}, while Yin et al. combined them with decision tree models \citep{25}.} Kolm et al. conducted an analysis of various deep-learning methods to forecast high-frequency returns across multiple horizons. They utilized order book information for 115 stocks traded on \textcolor{black}{the} Nasdaq. They compared the performance of models trained on order flow data versus those trained directly on the original limit order book \citep{26}, which has \textcolor{black}{significantly influenced our approach}.

The majority of research in this field has predominantly \textcolor{black}{focused on the U.S. and European market \citep{2}}. However, there has been a growing interest, particularly in the Chinese market \citep{3}, which has gradually garnered more attention in recent times. In the Chinese A-share market, the exchange publish the level-II LOB data (10 tiers) every three seconds, with 4500–5000 daily ticks \citep{4}. In the A-share market, the longer tick-time interval provides the opportunity to predict market movements and make profits by analyzing the LOB data. In this paper, we focused on the application of deep learning in the A-share market.

A lot of past research has obtained inspiration from other fields where the applications of deep learning models are mature, such as Natural Language Processing (NLP) and Computer Vision (CV), and has made satisfactory progress. However, these methods usually neglected to account for the unique characteristics and intricacies inherent in Limit Order Book (LOB) data. \textcolor{black}{For example, while techniques like LSTM and GRU have been adapted to process the temporal characteristics of LOB data, they often fail to fully exploit the specific properties that distinguish LOB data from other types of series data.} It is proven that in the U.S. equity market, the order flow imbalance (OFI) features transformed from the original LOB data can enhance the stability of the series and improve performance \citep{26,27,28}. \textcolor{black}{This indicates that certain transformations or representations of LOB data can lead to better modeling outcomes.} Inspired by this research, we recognized that an effective method for modeling LOB data should \textcolor{black}{align with its inherent characteristics to model the LOB more accurately}. We found that one of its most special traits is that the data structure has a strong symmetry between the ask and bid sides. \textcolor{black}{This symmetry is a fundamental property of LOB data, as the bid and ask sides represent the two sides of the market and are inherently mirror images of each other in terms of their structure and function.} However, few studies have explored the potential utilization of this unique feature. In this paper, we thoroughly analyzed the symmetry traits of LOB data and innovatively proposed a simple and effective Siamese architecture to extract features from the LOB data. \textcolor{black}{The core idea behind this architecture is to leverage the symmetry between the bid and ask sides to improve the modeling process.} We leveraged a parameter-sharing mechanism to keep the symmetry of the feature extraction process from both the buying and selling sides of the data \textcolor{black}{and improve the efficiency of data utilization.} \textcolor{black}{By using identical neural network structures for both sides and sharing parameters, we ensure that the model treats the bid and ask sides in a consistent manner, which is in line with their symmetric nature in the LOB data. This not only reduces the number of parameters and computational complexity but also enables the model to better capture the common patterns and relationships between the two sides.} This is complementary to existing methods, \textcolor{black}{which typically treat the bid and ask sides together and do not fully exploit their symmetry}. \textcolor{black}{Our Siamese architecture can be adapted to different model architectures, such as multi-layer perceptrons (MLPs), convolutional neural networks (CNNs), or recurrent neural networks (RNNs), making it a versatile and flexible solution for LOB data modeling, ultimately improving the performance of trading strategies and market analysis.}

Since the Multi-Head Attention (MHA) mechanism has achieved remarkable success across diverse complex sequence modeling tasks \citep{29}, its role \textcolor{black}{and potential impact} in LOB modeling remain unclear. Kolm et al. have conducted \textcolor{black}{comprehensive studies to} examine the predictive performance of various network structures, \textcolor{black}{including MLP, CNNs, and some RNNs,} for forecasting Nasdaq component stocks \citep{26}. However, their analysis did not consider the potential contribution of the attention mechanism, \textcolor{black}{which has shown great power in capturing long-term dependencies and informative features in other domains.} Given the condition, we incorporated both the LSTM and MHA mechanisms to examine the role of MHA in the stock price forecasting task. \textcolor{black}{The LSTM is capable of capturing temporal dependencies in the sequential stock data, and the attention mechanism can help focus on the most relevant information within the input sequences.} 

The contributions of this paper are summarized as follows:
\begin{itemize}
    \item First, we explored the symmetry traits of LOB data and proposed a simple and efficient Siamese architecture with parameter sharing to extract features for forecasting. Our method can be used as a supplement to the model structure and has good \textcolor{black}{generalization capabilities}.
    \item Then, we combined the MHA mechanism with the LSTM module to study the role of MHA on the forecasting results and discovered that MHA might help improve model performance over a small forecasting horizon.
    \item Finally, we have done experiments in the A-share market and proved that our method might improve the performances of different baselines in the A-share market except for the worst MLP, no matter the original LOB or OFI features as inputs. We also discussed the performance comparison between different types of inputs and stocks.
\end{itemize}

The outline of the article is as follows: In Section 2, we reviewed the basic knowledge of limit order book markets and \textcolor{black}{specified} the processed features used and related deep learning models. Then, we described our data, method, forecasting model, and evaluation methodology in Section 3. We presented the detailed experimental results and corresponding findings in Section 4. In Section 5, we summarized the conclusions and \textcolor{black}{proposed} the future prospects.

\section{Background and Related Work}
\subsection{The Limit Order Book}
\begin{figure}[h]
    \centering
    \includegraphics[width=1.0\linewidth]{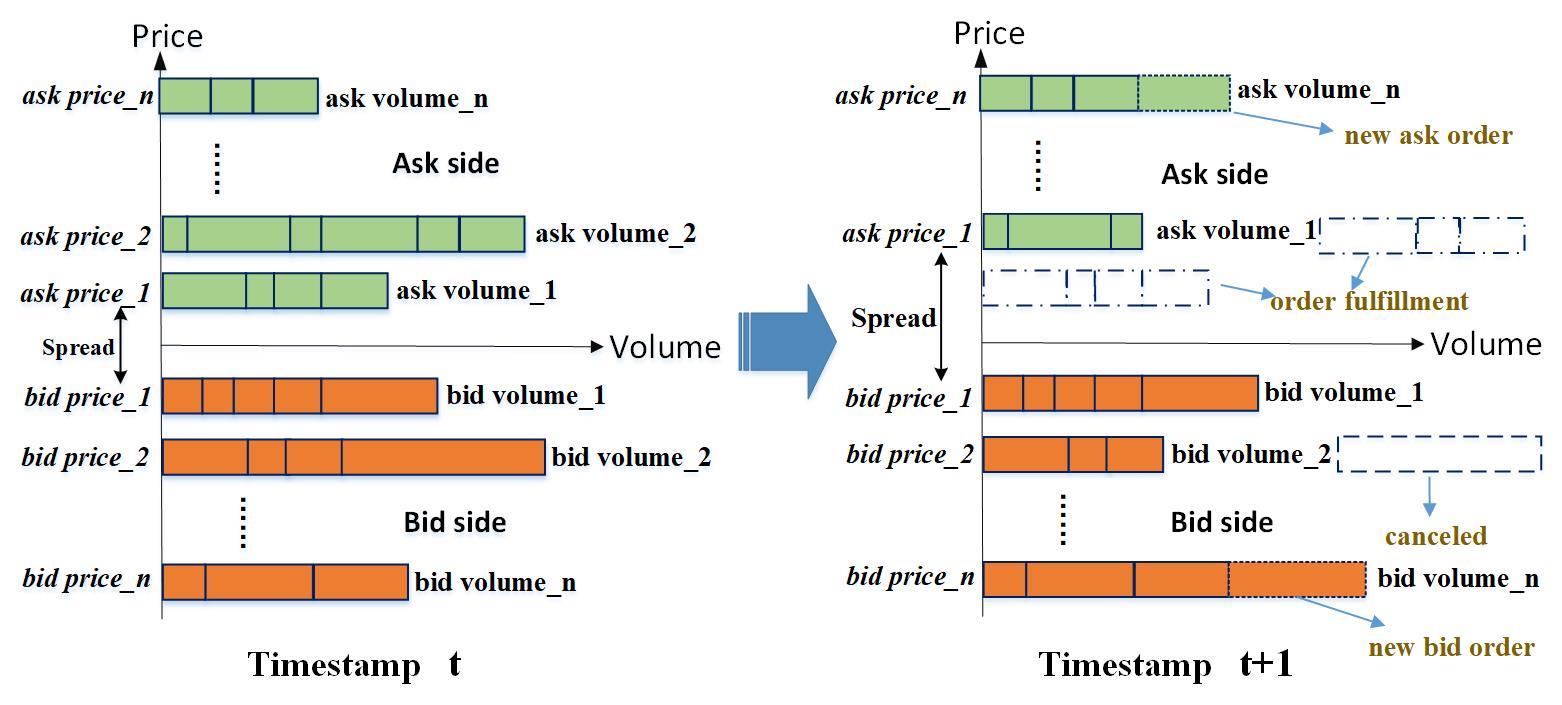}
    \caption{\centering \small The limit order book data structure.}
    \label{LOB_Data}
\end{figure}

 The Chinese A-share market adopts the continuous double auction mechanism to determine asset prices. The description "continuous" implies that traders can submit or withdraw their orders at any given moment during market hours. In the real stock market, buyers and sellers interact by submitting bid orders (buy) and ask orders (sell), which contain price and volume information. 
 The LOB data encompasses all the ask and bid prices along with their respective volumes.
 The bid orders are sorted in descending order of order prices, while the ask orders are sorted in ascending order, so the highest bid price and the lowest ask price are called "\textbf{best bid}" and "\textbf{best ask}", and the average of them is \textbf{"mid-price"}. The order execution rule is price priority (higher bid and lower ask first), then time priority (first come, first served). If traders choose to execute a certain number of buy/sell orders immediately at the current best price, regardless of the limit price, this type of order is called a \textbf{"market order"}. When a bid order price exceeds an ask order price, the stock exchange will match the corresponding orders into transactions directly, and the structure of LOB will change. Also, the arrival of new orders and the cancellation of old ones will change the LOB structure. In the A-share market, the exchange publishes the current status of LOBs every three seconds. The associated process is shown in Figure \ref{LOB_Data}.

\subsection{Deep Learning and Corresponding Feature Processing Methods}
In the past, predicting asset prices often required a mass of manually engineered features due to the high-dimensional and dynamic nature of LOB data. The quality of these features had a significant impact on the accuracy of the predictions. In recent years, researchers have begun to use deep learning techniques to predict stock price changes from LOB data. This has streamlined the process, reduced the burden of feature engineering, and enhanced predictive performance.

\textcolor{black}{Given that Huang et al. introduced LOBSTER, which is designed to process NASDAQ's ITCH data to accurately replicate the limit order book for any NASDAQ-traded stock, subsequent research has primarily focused on the NASDAQ market \cite{39}. } As the earliest application of neural networks to LOB modeling, Sirignano et al. proved that ordinary neural networks perform significantly better relative to logistic regression for modeling the distribution of the best ask and best bid prices without feature engineering \cite{30}. Also, they developed a novel spatial neural network architecture for modeling spatial distributions with lower computational expense and better generalization compared with common architectures. Then, Tsantekidis et al. proposed to utilize the CNN architecture to predict stock price movements as a classification task, and the results proved that this methodology is more effective than MLP and SVM methods \cite{17}. CNN can determine the market micro-structure to detect mid-price changes that occur. At the same time, as a structure suited well to sequence modeling, RNN was considered to model the regularity of LOB data. Dixon et al. utilize RNN to predict a next event price-flip from a short sequence of observations of LOB depths and market orders, and the results demonstrate that RNN can capture the non-linear relationship between the near-term price-flips and a spatiotemporal representation of the limit order book, so it compares favorably with other classifier methods, like the linear Kalman filter \cite{19}. Building on the successes of RNN and CNN, subsequent research has explored the potential of hybrid architectures that combine the strengths of these two architectures. Zhang et al. developed a deep-learning model that utilizes convolutional filters to capture the spatial structure of the LOB and LSTM modules to capture long-time dependencies \cite{18}. The results on the FI-2010 benchmark dataset show that this method performs better than other techniques in predicting short-term price movements and generalizes well to data that did not form part of the training data. Yin et al. utilized a similar CNN-LSTM architecture to the A-share market to design an expert trading system to integrate price prediction, trading signal generation, and optimization for capital allocation \cite{22}. The simulation demonstrates the trading system can make significant profits in different market sentiments after considering transaction costs and risks. 

In addition to the original LOB data, the OFI features derived from the LOB data are getting more and more attention due to their efficiency and simplicity. To explain price formation in LOB data and easily measurable inputs in simple ways, Cont et al. proposed a simple method to calculate OFI, and experimental results concluded that this simple linear relationship provides a strong link between order flow and price formation \cite{21}. Inspired by the work, Shen et al. and Xu et al. proposed to utilize the OFI features to fit the ordinary least squares (OLS) regression model to predict future price changes on Nasdaq, and they studied the relationship between the net order flow at the top price levels on each side and the concurrent movements in mid-price \cite{31,32}. The results uncover strong sample correlations between the net order flow at different price levels. Also, Kolm et al. combined the OFI features with various deep-learning methods for 115 stocks traded on Nasdaq, like MLP, LSTM, LSTM-MLP, and CNN-LSTM \cite{26}. From the results, models trained on order flow significantly outperformed most models trained directly on order books, proving the effectiveness of OFI compared with the original data on the asset price prediction tasks.

\textcolor{black}{Prata et al. examined the robustness and generalizability of 15 deep learning models for forecasting using LOB data, developing an open-source framework called LOBCAST for data preprocessing, model training, evaluation, and profit analysis. Extensive experiments reveal significant performance drops in all models when exposed to new data \cite{43}. Arroyo et al. proposed a deep learning method using a Convolutional-Transformer encoder and a monotonic neural network decoder to estimate limit order fill times in a limit order book (LOB), significantly outperforming traditional survival analysis approaches \cite{36}. Lucchese et al. employed deep learning techniques to conduct a large-scale analysis of predictability in high-frequency returns driven by order books, introducing a volume representation of the order book and conducting extensive empirical experiments \cite{37}. Jaddu et al. focused on forecasting returns across multiple horizons using order flow imbalance and training three temporal-difference learning models for five financial instruments, including forex pairs, indices, and a commodity \cite{40}. Zhang et al. proposed PAM-ENet, an ensemble network that combines a convolutional neural network (CNN) and two gated recurrent unit (GRU) networks to handle different types of LOB features, and uses a position attention mechanism (PAM) based fusion module to integrate their outputs \cite{41}. Kumar et al. presented a high-frequency market-making strategy that uses the Deep Hawkes process to create a feedback loop between order arrivals and the limit order book state, enabling agents to optimize pricing, order types, and execution timing \cite{42}. Briola et al. introduced "HLOB", a novel deep learning model for forecasting Limit Order Book mid-price changes, which leverages an Information Filtering Network and draws inspiration from Homological Convolutional Neural Networks to handle system complexity \cite{38}. These methods reveal the predictability in mid-price returns is widespread at high frequencies, and introducing the proper deep model might improve the forecasting performance.}

\section{Data, Feature, and Methodology}
In the past, most researchers focused on mature equity markets such as the U.S. or Europe. 
In this paper, our primary objective is to predict stock price changes within the A-share stock market. This section will introduce the research objects and outline our proposed methods.

\subsection{The LOB data and OFI features} \label{section_3.1}
In this paper, we \textcolor{black}{regarded} the level II LOB data (the top 10 trading prices and the corresponding volumes on each side) \textcolor{black}{as the research object, which is published by the exchanges and update every three seconds,} so there are about 4500–5000 ticks daily. We defined the state of the order book at the $j_{th}$ tick of trading day $i$ as the vector containing the top ten ask and bid tiers information as follows.

\begin{equation}
    s_{i,j}^{lob} := (a_{i,j}^1, v_{i,j}^{1,a},b_{i,j}^1, v_{i,j}^{1,b},...,a_{i,j}^{10}, v_{i,j}^{10,a},b_{i,j}^{10}, v_{i,j}^{10,b})^\top \in \mathbb{R}^{40} 
\end{equation}
where $a_{i,j}^k$, $b_{i,j}^k$ represent the ask and bid prices at $k_{th}$ tier at the $j_{th}$ tick of trading day $i$; $v_{i,j}^{k,a}$, $b_{i,j}^{k,b}$ represent the corresponding trading volumes.

Moreover, we can obtain the OFI features by applying a simple transformation to the LOB data, and the stability of the processed OFI distribution over time is significantly improved compared to the original LOB data distribution. Given two consecutive order book states for the stock at \textit{t-1} and \textit{t}, we can obtain the \textit{\textbf{bid order flows (bOF)}} and \textit{\textbf{ask order flows (aOF)}} at time \textit{t} as the representation $bOF_t \in \mathbb{R}^{10}$ and $aOF_t \in \mathbb{R}^{10}$
\begin{align} \label{ofi_equation}
    bOF_{t,i} :=& \begin{cases}
    v_{t}^{i,b}, & \text{ if } b_t^i > b_{t-1}^i,\\
    v_{t}^{i,b}-v_{t-1}^{i,b}, & \text{ if } b_t^i = b_{t-1}^i \\
    -v_{t}^{i,b}, & \text{ if } b_t^i < b_{t-1}^i
    \end{cases} 
    \\
    aOF_{t,i} :=& \begin{cases}
    -v_{t}^{i,a}, & \text{ if } a_t^i > a_{t-1}^i,\\
    v_{t}^{i,a}-v_{t-1}^{i,a}, & \text{ if } a_t^i = a_{t-1}^i \\
    v_{t}^{i,a}, & \text{ if } a_t^i < a_{t-1}^i
    \end{cases}
\end{align}
where i = 1,...,10. The nonlinear transformations have become a common approach to converting the nonstationary time series of order book states to stationary ones \citep{21}. By concatenating these two components, we obtained the whole order flow imbalance (OFI) state $s_t^{ofi}$ at time t 
\begin{equation}
    s_t^{ofi} := \left (  \begin{matrix}
        bOF_t \\   aOF_t
  \end{matrix}  \right ) \in \mathbb{R}^{20}
\end{equation}
 
The utilization of the OFI feature treats the contributions of the market-, limit-, and cancel-orders equally \citep{21}, and experimental results demonstrate that OFI performs better than the original LOB in the Nasdaq market \citep{31,32,33}. In the following, I will separately study the role of these two types of features in the A-share market.

\subsection{Dataset and Evaluation}
We conducted a research study on the following 14 defense-industry-related stocks with good liquidity, from January 6, 2021, to May 13, 2021, and all 14 stocks are listed on the A-share market:

\noindent \hrulefill

\noindent 
\small 003026.SZ, 300864.SZ, 300870.SZ, 300877.SZ, 300881.SZ, 300886.SZ, 300892.SZ, 

\noindent 
\small 300896.SZ, 300898.SZ, 300908.SZ, 300910.SZ, 300919.SZ, 300925.SZ, 300999.SZ.

\noindent \hrulefill 

\begin{figure}[htb]
    \centering
    \includegraphics[width=0.90\linewidth]{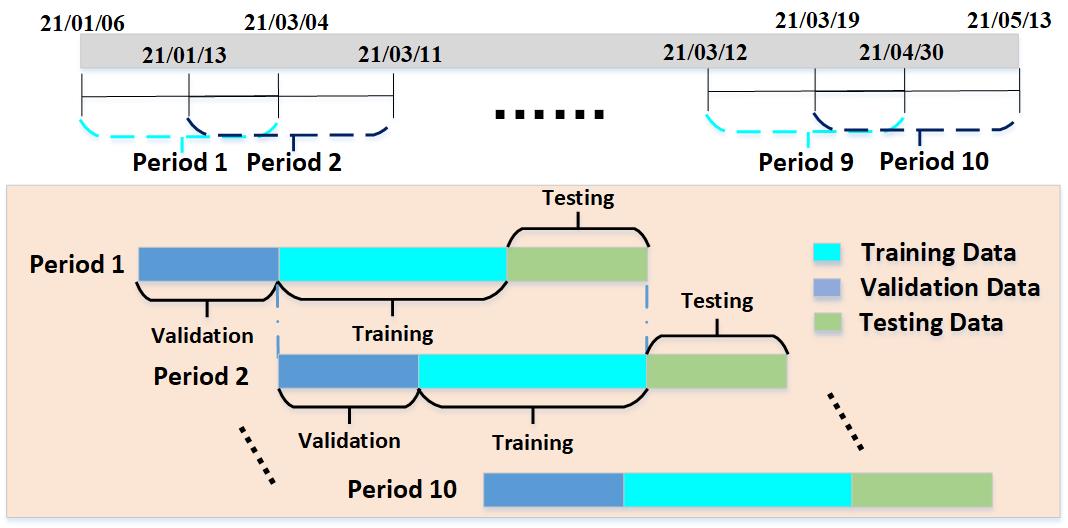}
    \caption{\centering \small The train/valid/test sets split.}
    \label{dataset_split}
\end{figure}

Since the price movement patterns of stocks are affected by the market and do not obey the assumption of \textcolor{black}{independent and identically distributed (i.i.d.)} distribution over a long period, we considered a rolling-window backtesting fashion to fit the real-world production setting, as used in \cite{26}. The first week is used for validation, the following five weeks are used for training, and the last week is reversed for out-of-sample testing. The specific procedure is shown in Figure \ref{dataset_split}.

Like most studies, we employed the observed historical trading information to predict asset price changes. For the trading tick t, we denoted the input data as a sequence of vectors $X_t=[s_{t-\delta},...,s_t] ^\top \in \mathbb{R}^{(\delta +1) \times 10d}$, where $\delta$ \textcolor{black}{is} the length of historical tick information. For LOB, d=4; and for OFI, d=2. We set $\delta$ to 49 in our experiment, which means that 50 historical ticks (about 150 sec) are used, and this length is appropriate in the A-share market \citep{4}.

We have identified distribution shifts in our trading data due to the overnight stock price jump. The most common \textcolor{black}{normalization} methods, such as z-score and min-max \citep{3}, are hard to overcome this problem. To deal with this problem, we subtracted the closing price of the previous trading day from the asset price information. This simple method \textcolor{black}{was} proposed in \cite{33} as a part of the NLinear method and performs well in many time-series forecasting tasks.

In this paper, we formulated the price forecasting task as a regression problem to model asset price movements more clearly. Our objective is to predict the mid-price changes \textit{in the subsequent h ticks time interval} (\textbf{h is the forecasting horizon}) compared to the observed mid-price at the current. For trading time k, the target price change is expressed as follows:
\begin{equation}
    r_k = \frac{1}{h}\sum _{t=k+1}^{t=k+h}\frac{(a_t^1+b_t^1)}{2} - \frac{(a_k^1+b_k^1)}{2}
\end{equation}

In most cases, stock price changes are gradual, typically at 0.01 China Yuan (CNY), but there are occasional price jumps, \textcolor{black}{which always might be considerd as anomalies}. To mitigate the impact of these anomalies, we capped the absolute value of the target at 1 CNY, \textcolor{black}{so the model can focus on learning the normal price changes and reduce the influence of these rare large jumps}. If the price change exceeds the threshold, for example, if the real price increase is 1.5 CNY, the price tag is set to 1 CNY. \textcolor{black}{We think this approach can help improve the stability and accuracy of the model.}

As a regression task, we first measured the forecasting performance using the mean absolute error (MAE) on the test set. MAE, widely used in regression tasks, represents the average \textcolor{black}{absolute} difference between actual and predicted price changes. Assuming that the price change at time t is $p_i$ and the predicted result is $\hat{p_i}$. Another similar metric is the mean-square error (MSE), MAE and MSE at horizon h are expressed as
\begin{align}
    MAE_h &= \sum_{i=1}^{N} \|p_i - \hat{p}_i\| \\
    MSE_h &= \sum_{i=1}^{N} (p_i - \hat{p}_i)^2,
\end{align}
where N represents the number of samples on the test set.
In addition, we adopted out-of-sample $R^2$ to assist in measuring the ability of the forecast model to explain price changes at horizon h ($R^2_{OS,h}$) for each test period, defined as

\begin{align}
    R^2_{OS,h} &= 1 - \frac{MSE_{m,h}}{MSE_{bmk,h}}  \\
    MSE_{bmk,h} =& \sum_{i=1}^{N} (p_i - \frac{1}{N}\sum_{k=1}^{N}p_k)^2
\end{align}
Where $MSE_{m,h}$ and $MSE_{bmk,h}$ are separately the MSE of the model forecasts ad benchmark at the $h_{th}$ horizon, and we utilized the average out-of-sample return as our benchmark. If $R^2_{OS,h}>0$, then the \textcolor{black}{predicted} model outperforms the benchmark proposed by the average return on the test set.

\subsection{Model and Training}

In this section, we focus on the model architectures used in our study. \textcolor{black}{We considered several architectures inspired by past research on the Nasdaq market, including MLP, stacked LSTM, MLP-LSTM, and CNN-LSTM.} Additionally, we investigated the role of the attention mechanism in this task and proposed an LSTM-MHA architecture that combines LSTM and Multi-Head Attention (MHA). The detailed model structures and parameters are provided in Appendix \ref{baselines}.

\begin{figure}[htb]
    \centering
    \includegraphics[width=0.80\linewidth]{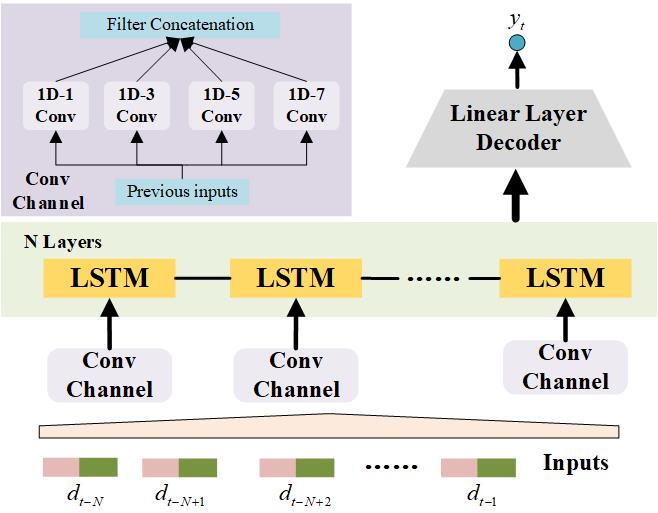}
    \caption{\centering \small The CNN-LSTM model architecture overview.}
    \label{cnn_lstm}
\end{figure}

The LSTM-MHA model we proposed consists of a stacked LSTM module and an MHA module. \textcolor{black}{Belonging to the family of RNNs, LSTM} alleviates the vanishing gradient problem through its unique gate mechanism, so it is widely used in various sequence modeling tasks. The essential elements of the LSTM are three gates that determine the information flow in, process, and out, referred to as input, forget, and output gates. Let $d_k \in \mathbb{R}^{N}$ denote an input vector of order k, the LSTM unit takes the form
\begin{align}
    f_k = & \sigma(U^fd_k + W^fh_{k-1} + b^f)   \\
    i_k = & \sigma(U^id_k + W^ih_{k-1} + b^i)    \\
    o_k = & \sigma(U^od_k + W^oh_{k-1} + b^o)    \\
    s_k =  &f_k\odot s_{k-1} + i_k\odot \tanh{(U^sd_k + W^sh_{k-1} + b^s)}   \\
    h_k = & o_t\odot \tanh{(s_k)}
\end{align}
where $\sigma:=(1+e^{-x})^{-1}$ represents the sigmoid activation function, $f_k \in \mathbb{R}^{D}$, $i_k \in \mathbb{R}^{D}$, $o_k \in \mathbb{R}^{D}$, and $h_k \in \mathbb{R}^{D}$ denote the forget gate's, input gate's, output gate's, and hidden state vector. $W^p \in \mathbb{R}^{D\times N}$, $U^p \in \mathbb{R}^{D\times N}$ and $b^p \in \mathbb{R}^{D} (p \in \{f,i,o\} )$ are trainable parameters. Here, $\odot$ and $tanh$\ represent the element-wise product and hyperbolic tangent operators.

\begin{figure}[htb]
    \centering
    \includegraphics[width=0.91\linewidth]{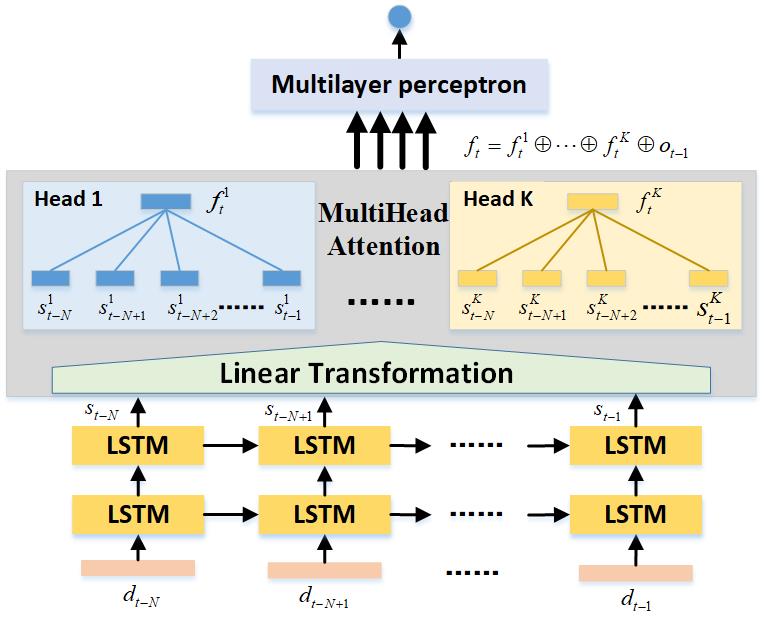}
    \caption{\centering \small The LSTM-MHA model architecture overview.}
    \label{attention_model}
\end{figure}

Then, the output from the stacked LSTM frequently feeds into the MHA architecture, whose role is to model long-distance dependencies between input time series. The idea of the attention mechanism is to compress the hidden states at different time steps into a total representation, and all time steps contribute differently to the final results. For predicting the price changes after time $t$, the state's series of LSTM cells are $[s_{t-N},...,s_{t-1}]$, and the final output is $o_{t-1}$. We utilized this architecture to decode the cell states of the LSTM module.
\begin{align}
    y_t =& W_of_t+b_o, \quad   f_t = f_t^1 \oplus ...\oplus f_t^K \oplus o_{t-1} \\
    f_t^k =& \sum_{m=t-N}^{t-1}  a_m^k \dot s_m^k, \quad  (k=1,...,K) \\
    a_m^k =& \frac{e^{\hat{a_m^k}}}{\sum_{n=t-N}^{t-1} a_n^k}  \\
    a_m^k = &o_{t-1}^\top \tanh(W_as_m^k +  b_a), \quad (s_m = s_m^1 \oplus ...\oplus s_m^K)
\end{align}
where $\oplus$ represents the concatenate operator. The head number of MHA, $K$, must be divisible by the dimension $D$. Assume $h= \tfrac{D}{K}$, $s_m^k \in \mathbb{R}^h$ and $o_{t-1} \in \mathbb{R}^D$, $W_a \in \mathbb{R}^{D\times h}$,  $W_o \in \mathbb{R}^{1\times 2D}$, $b_a \in \mathbb{R}^{D}$, and the scalar $b_o$  are trainable parameters, $\hat{y_t}$ is the forecasting result. In this paper, D and K are separately set to 64 and 4, and the overall model architecture is shown in Figure \ref{attention_model}.

\textcolor{black}{The work is inspired by previous research, and all baselines except LSTM-MHA have been studied, demonstrating the deep learning models work well, especially the CNN-LSTM model \cite{26}.} We also explored the use of transformer structures for price forecasting, but the results showed that both causal and non-causal transformers performed poorly, lagging behind all other models. \textcolor{black}{We believe it might be two reasons. Firstly, the Transformer model typically requires a large amount of data for effective training, and the LOB data we used may not be sufficient to fully train the model. Secondly, the Transformer architecture, which has achieved significant breakthroughs in the fields of Natural Language Processing (NLP) and Computer Vision (CV), usually requires the input data to be discretized \cite{ptst}, such as tokenization in NLP and patchify in CV. This is not well-suited to the continuous and dynamic nature of LOB data, and relevant conclusions were found in the time series forecasting tasks\cite{33}.}

We trained our models by minimizing \textcolor{black}{the MAE loss} with the Adam optimizer. \textcolor{black}{The hyperparameters we utilized is $\beta_1=0.9,\beta_2=0.999,eps=1e-8$, which is the default value in Pytorch.} The initial learning rate is set to 0.0001 and the weight decay is 0.001 \textcolor{black}{for regularizition} \citep{34}. The batch size in the training process is 256. We adopted an early stop to avoid overfitting, stopping training when \textcolor{black}{the MAE loss in the validation set} has not decreased for 5 consecutive epochs.

\textcolor{black}{All hyperparameters were selected through grid search method and informed by related work to ensure optimal performance. For instance, the learning rate was chosen from a range between 1e-3 and 1e-4, weight decay was tested at 0.01, and 0.001, and batch size was experimented with using 64, 256, and 512. Some default hyperparameters from the Adam optimizer were also utilized. The selection of hyperparameters took into account the scale of the data, the specific task, and the experimental conditions. For example, for a small training dataset, a large batch size is not suitable as it can lead to overfitting. The learning rate should not be too small, or the model will struggle to converge. Weight decay, which helps prevent overfitting through regularization, should not be set too large for tasks that are not prone to overfitting, as this can negatively impact predictive performance.}

All the experiments are carried out on a server equipped with GPU Tesla P100 with 12GB RAM onboard, Intel Xeon CPU E5-2680 v4 Processor, 2.40GHz, 128.0GB RAM, and the deep learning framework used is Pytorch. 

\subsection{\textcolor{black}{The Siamese architecture}}

LOB data is highly dimensional, noisy, and time-varying, making it difficult to capture patterns of price movements directly from the data. The motivation for our work is to effectively utilize the characteristics of the LOB structure to accurately forecast the stock price. We \textcolor{black}{assume} that one of the most significant traits of the LOB data is its symmetry. Specifically, we can divide the LOB data into buy-side and sell-side, and the data \textcolor{black}{structures} on both sides are identical. The price and volume range from \textit{ask/bid 1} to \textit{ask/bid N}. The order of transactions \textcolor{black}{gives} priority to ask/bid 1, which is the best bid/ask price, and continues in turn. This ensures that the overall LOB is \textcolor{black}{symmetrical} around the mid-price. Intuitively, the symmetry inherent in LOB data is \textcolor{black}{advantageous} for modeling. For instance, the calculation of $aOF$ and $bOF$ features mentioned in Section \ref{section_3.1} \textcolor{black}{is} also highly symmetrical, implicitly employing the symmetry traits of LOB data. 

From a holistic perspective, all the baselines \textcolor{black}{comprise} two modules: a feature extraction encoder and a prediction decoder. The encoder part extracts relevant features from the original high-dimensional LOB data, which \textcolor{black}{combines} feature engineering and generates representations, and the commonly used methods are CNN, LSTM, etc. The decoder part \textcolor{black}{is dedicated to} forecasting future price movements based on features obtained from the encoder, which can be implemented using a simple MLP. Raw LOB data or processed features \textcolor{black}{were} directly sent to the encoder in the past. In this paper, we \textcolor{black}{propose} a new approach based on the Siamese network, taking full advantage of symmetry traits to enhance the performance of baselines. We utilized two encoders to obtain the ask- and bid-related representations ($f_t^a$ and $f_t^b$) from the ask- and bid-side data, respectively. Considering the symmetry of the two sides, the two encoders have the same structure and share parameters, which improves the training efficiency and \textcolor{black}{ensures} the symmetry of obtained features. This structure, known as a Siamese network, has been used to compare \textcolor{black}{similarity} between two inputs for tasks such as text matching \citep{35}. To the best of our knowledge, we are the first to use this structure for LOB data, and it is instructive for similar task forms. The model framework \textcolor{black}{differences} between the two forms is shown in Figure \ref{siamesed_model}. The encoder can adopt various structures like CNN, LSTM, etc., and our approach scales well across various architectures.

\begin{figure}[htb]
    \centering
    \includegraphics[width=1.0\linewidth]{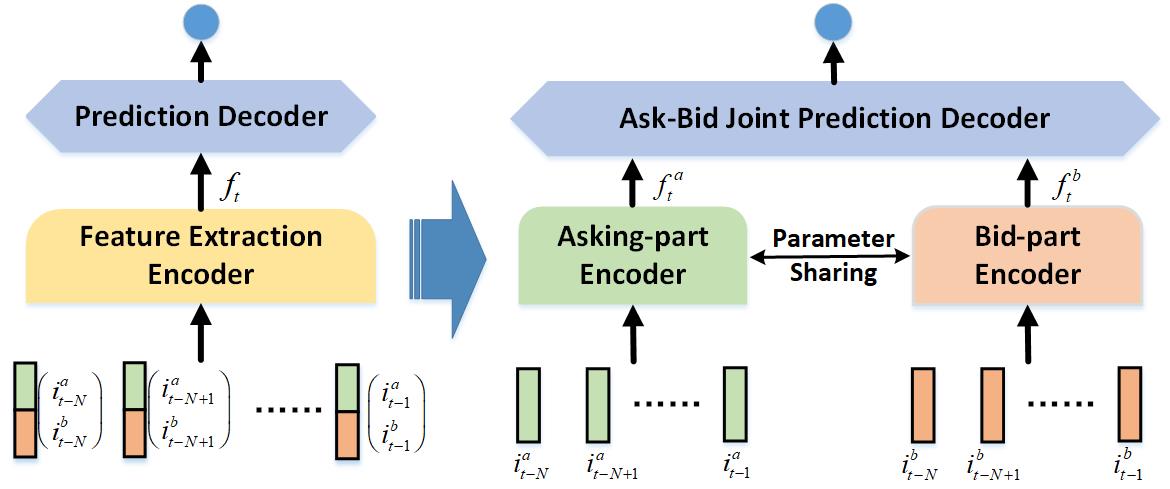}
    \caption{\centering \small The Siamese network architecture.}
    \label{siamesed_model}
\end{figure}

In addition, as we \textcolor{black}{get} both $f_t^a$ and $f_t^b$ feature representations corresponding to ask and bid sides, we \textcolor{black}{consider} adjusting the prediction decoder. The simplest method is to concatenate two features as input of the decoder, and we choose to feed the subtraction of the two features into a 2-layer MLP to predict the result. Since the expected price \textcolor{black}{changes} are not beyound 1 CNY, the sigmoid activation function $\sigma$ \textcolor{black}{is} used to ensure the range of the output.
\begin{align}
    \hat{y_t} = \sigma(W_o^2(W_o^1(f_t^a - f_t^b) + b_o^1) + b_o^2) \times \hat{\alpha} - \hat{\beta}
\end{align}
where $W_o^1 \in \mathbb{R}^{D_2\times D}$, $W_o^2 \in \mathbb{R}^{1\times D_2}$, $b_o^1 \in \mathbb{R}^{D_2}$  and scalar $b_o^2$ are trainable parameters, and the scalar $\hat{y_t}$ represents the predicted result. $\hat{\alpha}$ and $\hat{\beta}$ are used to align the output with the range of actual price movements.

\section{Experiment Results and Discussion}
In this section, we summarized our experimental results and discuss them. First, in Section \ref{section 4.1}, we showed performance differences on a baseline basis using our proposed Siamese network framework, \textcolor{black}{and} further demonstrated the factors that influence the performance of baselines and their impact. The factors affecting performance are further discussed in Section \ref{section 4.3}.

\subsection{\textcolor{black}{Main Results and Analysis}} \label{section 4.1}

\begin{figure}[htb]
    \centering
    \ContinuedFloat
    \subfigure[\footnotesize MLP (MAE).]{
      \begin{minipage}[t]{0.43\textwidth}
        \includegraphics[width=\textwidth]{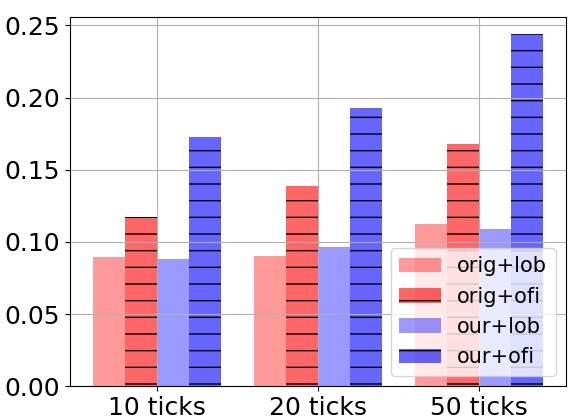}
        \label{mlp_mae}
      \end{minipage}
      }
  \hfill
  \subfigure[\footnotesize MLP ($R^2$).]{
      \begin{minipage}[t]{0.45\textwidth}
        \includegraphics[width=\textwidth]{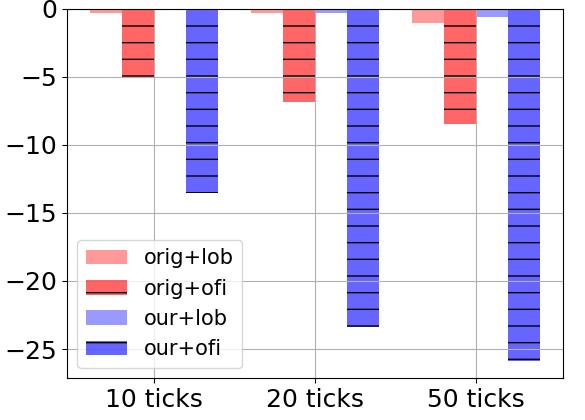}
        \label{mlp_r2}
      \end{minipage}
      }
  \hfill
  \subfigure[\footnotesize LSTM (MAE).]{
      \begin{minipage}[t]{0.45\textwidth}
        \includegraphics[width=\textwidth]{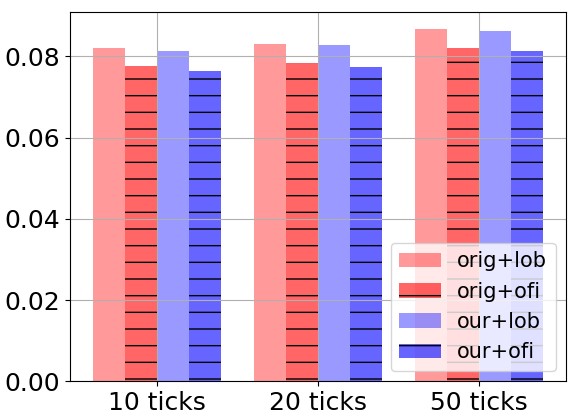}
        \label{tab4_c}
      \end{minipage}
      }
  \hfill
  \subfigure[\footnotesize LSTM ($R^2$).]{
      \begin{minipage}[t]{0.45\textwidth}
        \includegraphics[width=\textwidth]{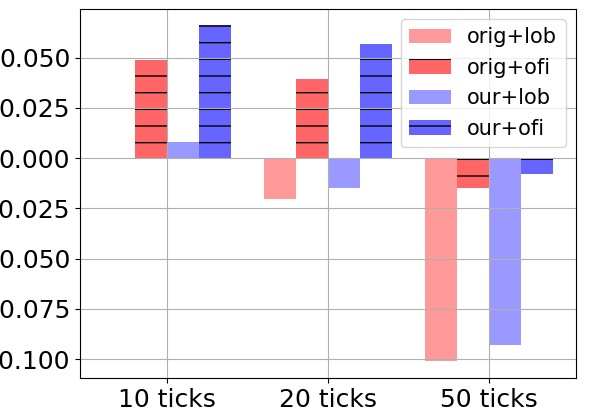}
        \label{tab4_d}
      \end{minipage}
      }
    \hfill
    \subfigure[\footnotesize MLP-LSTM (MAE).]{
      \begin{minipage}[t]{0.45\textwidth}
        \includegraphics[width=\textwidth]{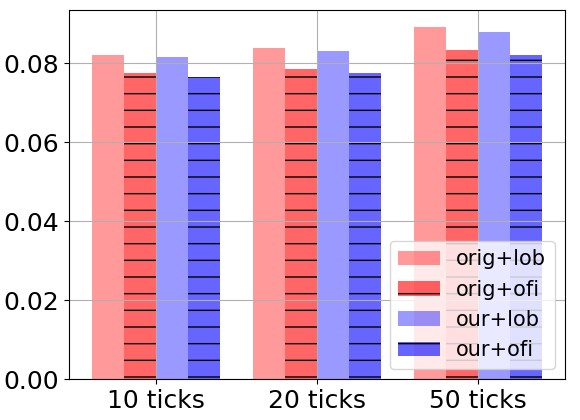}
        \label{mlp_mae}
      \end{minipage}
      }
  \hfill
  \subfigure[\footnotesize MLP-LSTM ($R^2$).]{
      \begin{minipage}[t]{0.45\textwidth}
        \includegraphics[width=\textwidth]{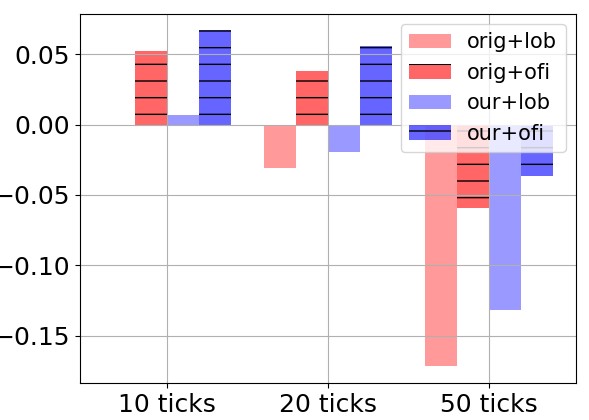}
        \label{mlp_r2}
      \end{minipage}
      }
\end{figure}
\begin{figure}
    \ContinuedFloat
  \hfill
  \subfigure[\footnotesize CNN-LSTM (MAE).]{
      \begin{minipage}[t]{0.45\textwidth}
        \includegraphics[width=\textwidth]{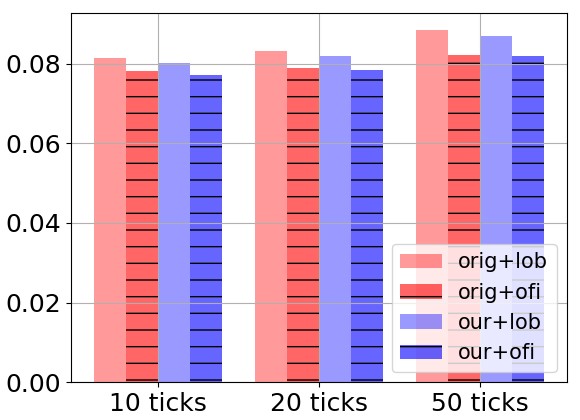}
        \label{cnn_lstm_mae}
      \end{minipage}
      }
  \hfill
  \subfigure[\footnotesize CNN-LSTM ($R^2$).]{
      \begin{minipage}[t]{0.45\textwidth}
        \includegraphics[width=\textwidth]{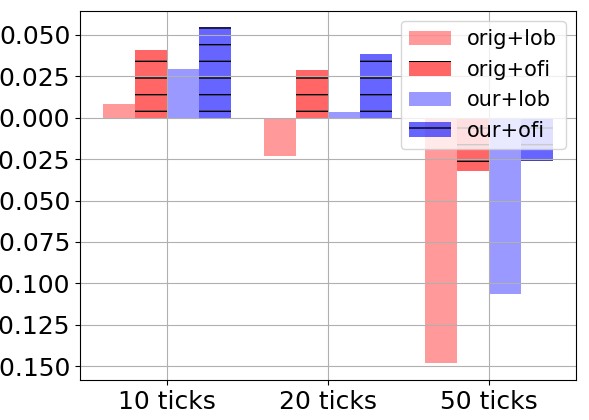}
        \label{cnn_lstm_r2}
      \end{minipage}
      }
    \hfill
    \subfigure[\footnotesize LSTM-MHA (MAE).]{
      \begin{minipage}[t]{0.45\textwidth}
        \includegraphics[width=\textwidth]{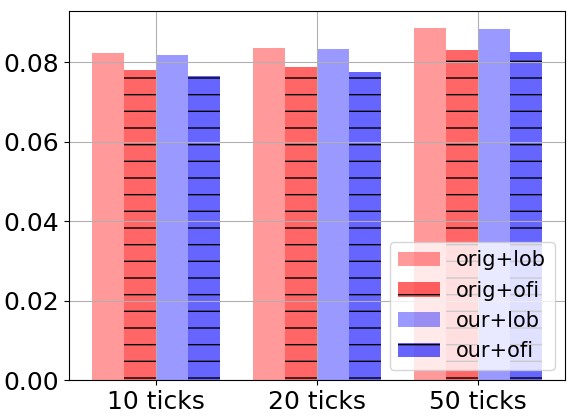}
        \label{attn_mae}
      \end{minipage}
      }
  \hfill
  \subfigure[\footnotesize LSTM-MHA ($R^2$).]{
      \begin{minipage}[t]{0.45\textwidth}
        \includegraphics[width=\textwidth]{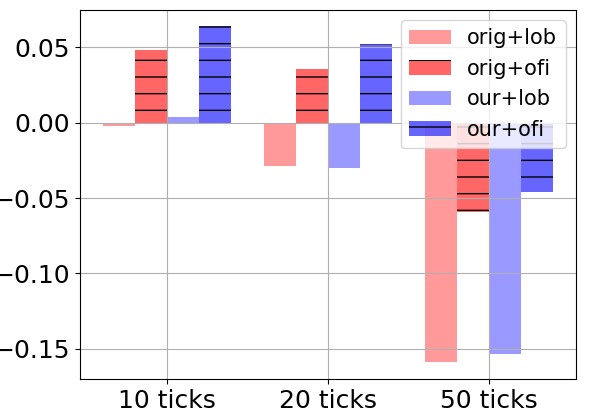}
        \label{attn_r2}
      \end{minipage}
      }
    \hfill
    \caption{\centering \small The performance results of all baselines on different horizons.}
    \label{results}
\end{figure}

Figure \ref{results} shows the \textcolor{black}{performance comparision} of the original baseline versus the proposed Siamese architecture. \textcolor{black}{The results are the MAE and $R^2$} at horizons h = 10, 20, and 50 ticks, \textcolor{black}{which reflected the ability to predict price movements after 0.5, 1, and 2.5 minutes.} The smaller the MSE, the larger the $R^2$, and the better the prediction performance and model capability. "orig+" represented the results of original baselines, and "our+" represented the results when utilizing \textcolor{black}{the methods proposed in the manuscript} on baselines.

It can be found that with the extension of the predicted horizon range, the prediction effect of the model gradually deteriorates, and the result is consistent with our intuition because of the uncertainty of future transaction events. \textcolor{black}{As the prediction horizon increases, the number of potential transaction events and their associated uncertainties also increase, making it more challenging for the model to accurately forecast future price movements.}

After the Siamese network framework is adopted, the MSE and $R^2$ performances of baselines are significantly improved, especially for $R^2$. \textcolor{black}{The Siamese network leverages the symmetry traits of LOB data, which allows the model to better capture the inherent structure and relationships within the data. This symmetry-based approach enhances the model's ability to generalize and improves its predictive accuracy. The significant improvement in $R^2$ indicates that the model is better able to explain the variance in the target variable, which is crucial for forecasting tasks.}  No matter \textcolor{black}{whether} the original LOB or processed OFI \textcolor{black}{is used} as input, the proposed method can obtain a significant performance improvement. This indicates that even after feature processing, the symmetry nature can significantly promote performance improvement. \textcolor{black}{The symmetry in LOB data provides a structural advantage that allows the model to better capture underlying patterns, thereby enhancing its predictive capabilities.} We hope this will shed some light on better mining useful traits for price forecasting and modeling.

In addition to \textbf{MLP}, utilizing the OFI features performed better than using raw LOB data in the rest of the baselines, consistent with the results on the Nasdaq market \citep{26}. \textcolor{black}{The superior performance of OFI features can be attributed to their ability to capture more informative and discriminative patterns from the LOB data, which helps improve the model's predictive power. Specifically, OFI features are designed to reflect the relationship developments between supply and demand, which is the direct driver of price movements in financial markets.}  The result of \textbf{MLP} on the A-share market is obviously inferior to the rest of the baselines and may not be suitable for the A-share market directly. \textcolor{black}{The poor performance of MLP on the A-share market could be due to its limited capacity to capture complex patterns and dependencies in the data, leading to suboptimal performance.}

\textcolor{black}{To further analyze the role of OFI features for better performance}, we trained all ten models on the LOB or OFI inputs, \textcolor{black}{resulting in} twenty model and input combinations. To compare their out-of-sample performance on different test sets, we ranked them by their average forecast MSE as follows. For each stock, test set, and horizon, we ranked the combinations from best to worst based on their MSE results, such that the best combination receives a rank of 1 and the worst receives 20. Then, based on the methods of order measurement, we averaged the reciprocal of rank position on all the test sets as the global average score for each model, input, and horizon, like the formula (\ref{score}) shows. The overall result demonstrates in table \ref{table_1.1}, the model score and corresponding ranking (numbers in brackets).

\begin{align} \label{score}
    Score_{md}^{h} = \frac{1}{N}\sum_{k=1}^{N} \frac{1}{rank_{md}^{k,h}}
\end{align}
where $Score_{md}^{h}$ represents the score of model \textit{md} at horizon \textit{h}. \textbf{N} is the total number of test sets. In this paper, we have 14 stocks, and each stock contains 10 or 11 test sets, and $N=149$. $rank_{md}^{k,h}$ is the performance rank of model \textit{md} at test set \textit{k}, horizon \textit{h}, range from 1 to 20.

\begin{table}[h]
  \centering
  \caption{The performance comparison of different methods on different horizons.}
    \begin{tabularx}{\textwidth}{XXX|XXXXX}
    \hline
    \multirow{2}{*}{Horizon} & \multirow{2}{*}{Feature} & \multirow{2}{*}{Method} & \multicolumn{5}{c}{Model Type} \\
          &       &       & MLP   & LSTM  & MLP-LSTM & CNN-LSTM & LSTM-MHA \\
    \hline
    \multirow{4}{*}{10} & \multirow{2}{*}{LOB} & Original & 0.064(20) & 0.069(18) & 0.071(17) & 0.076(16) & 0.068(19) \\
          &       & Siamese & 0.107(12) & 0.080(15) & 0.086(13) & 0.122(9) & 0.082(14) \\
          & \multirow{2}{*}{OFI} & Original & 0.116(10) & 0.177(7) & 0.191(5) & 0.107(11) & 0.198(4) \\
          &       & Siamese & 0.126(8) & 0.506(3) & 0.552(2) & 0.183(6) & 0.615(1) \\
    \hline
    \multirow{4}{*}{20} & \multirow{2}{*}{LOB} & Original & 0.068(19) & 0.078(15) & 0.067(20) & 0.073(18) & 0.089(13) \\
          &       & Siamese & 0.110(9) & 0.093(12) & 0.075(17) & 0.098(11) & 0.078(16) \\
          & \multirow{2}{*}{OFI} & Original & 0.082(14) & 0.219(5) & 0.185(7) & 0.239(8) & 0.229(4) \\
          &       & Siamese & 0.105(10) & 0.537(2) & 0.499(3) & 0.185(6) & 0.596(1) \\
    \hline
    \multirow{4}{*}{50} & \multirow{2}{*}{LOB} & Original & 0.071(18) & 0.090(16) & 0.064(20) & 0.070(19) & 0.090(15) \\   &       & Siamese & 0.098(10) & 0.100(9) & 0.078(17) & 0.090(14) & 0.095(12) \\
          & \multirow{2}{*}{OFI} & Original & 0.097(11) & 0.378(3) & 0.147(8) & 0.165(7) & 0.266(5) \\
          &       & Siamese & 0.093(13) & 0.614(1) & 0.288(4) & 0.219(6) & 0.485(2)  \\
    \hline
    \end{tabularx}%
  \label{table_1.1}%
\end{table}%

It can be seen from the results that the results based on the OFI feature are significantly better than those based on original LOB data in model ranking. After our Siamese network framework was adopted, the model ranking performance improved, demonstrating the framework we proposed is effective. The performance of MLP-LSTM and CNN-LSTM models is not better than that of LSTM and even deteriorates when the OFI feature is used, being different from the Nasdaq market. We assumed it may be due to the low-frequency update of LOB in the A-share market. Unlike the millisecond update time on Nasdaq, LOB data on the A-share market updates every 3s, so processing the data before feeding it into the LSTM does not bring improvements. The LSTM-MHA we proposed performs inferior on LOB data but performs better than other baselines when using the OFI feature. We believed the MHA mechanism can further enhance the feature performance if the feature is good. But if the feature performance is poor, introducing the MHA mechanism may lead to performance degradation.

Finally, we analyzed the MAE performance of different test sets in detail, comparing the features and architectures. Tables \ref{table_1.2} and \ref{table_1.3} separately demonstrate the number of test sets with better results under various horizon conditions on the 149 test data. In addition to MLP, the OFI feature is significantly better than LOB data on different test sets, and the Siamese network framework is also better than the baseline, except for LSTM-MHA on LOB data. \textcolor{black}{The Siamese network framework, by leveraging the symmetry traits of LOB data, enhances the model's ability to generalize and improves its predictive accuracy. The LSTM-MHA model, while performing inferior on LOB data, shows better performance when using the OFI feature, indicating that the MHA mechanism can effectively enhance the feature's performance when the feature itself is strong. However, if the feature performance is poor, introducing the MHA mechanism may lead to performance degradation.}

\begin{table}[h]
   \centering
   \caption{\centering The performance comparison between original LOB and OFI features (the "win times" on all test sets) For example, at horizon 10, the MLP utilizing the  original LOB performs better on 60 test sets, and the OFI acts better on 89 test sets.}
    \footnotesize
    \begin{tabular}{cc|cc|cc}
    \hline
    \multirow{2}{*}{Horizon} & \multirow{2}{*}{Model} & \multicolumn{2}{c|}{Original Method} & \multicolumn{2}{c}{Siamese network} \\
         &       & LOB   & OFI   & LOB   & OFI \\
    \hline
    \multirow{5}{*}{10} & MLP   & 60    & 89    & 77    & 70 \\
          & LSTM  & 3     & 146   & 1     & 148 \\
          & MLP-LSTM & 3     & 145   & 3     & 146 \\
          & CNN-LSTM & 18    & 120   & 28    & 108 \\
          & LSTM-MHA & 3     & 146   & 2     & 147 \\
    \hline
    \multirow{5}{*}{20} & MLP   & 76    & 72    & 92    & 56 \\
          & LSTM  & 4     & 145   & 3     & 146 \\
         & MLP-LSTM & 4     & 145   & 1     & 148 \\
          & CNN-LSTM & 8     & 140   & 9     & 135 \\
          & LSTM-MHA & 4     & 145   & 4     & 145 \\
    \hline
    \multirow{5}{*}{50} & MLP   & 76    & 73    & 96    & 53 \\
          & LSTM  & 5     & 143   & 7     & 142 \\
         & MLP-LSTM & 4     & 145   & 5     & 144 \\
          & CNN-LSTM & 5     & 144   & 3     & 144 \\
          & LSTM-MHA & 2     & 147   & 4     & 145 \\
    \hline
    \end{tabular}%
  \label{table_1.2}%
\end{table}%

\begin{table}[h]
  \centering
  \caption{The performance comparison between original and Siamese network.}
    \footnotesize
    \begin{tabular}{cc|cc|cc}
    \hline
    \multirow{2}{*}{Horizon} & \multirow{2}{*}{Model} & \multicolumn{2}{c|}{LOB Feature} & \multicolumn{2}{c}{OFI Feature} \\     &       & Original  & Siamese & Original  & Siamese \\
    \hline
    \multirow{5}{*}{10} & MLP   & 34    & 106   & 72    & 71 \\
          & LSTM  & 29    & 107   & 2     & 142 \\
          & MLP-LSTM & 34    & 99    & 9     & 132 \\
          & CNN-LSTM & 5     & 137   & 9     & 134 \\
          & LSTM-MHA & 37    & 102   & 2     & 128 \\
    \hline
    \multirow{5}{*}{20} & MLP   & 35    & 112   & 75    & 68 \\
          & LSTM  & 43    & 98    & 7     & 136 \\
          & MLP-LSTM & 33    & 110   & 8     & 131 \\
          & CNN-LSTM & 15    & 130   & 21    & 115 \\
          & LSTM-MHA & 62    & 72    & 14    & 130 \\
    \hline
    \multirow{5}{*}{50} & MLP   & 22    & 127   & 93    & 56 \\
          & LSTM  & 54    & 90    & 33    & 108 \\
          & MLP-LSTM & 48    & 92    & 20    & 124 \\
          & CNN-LSTM & 25    & 121   & 35    & 101 \\
          & LSTM-MHA & 80    & 60    & 37    & 106 \\
    \hline
    \end{tabular}%
  \label{table_1.3}%
\end{table}%

\subsection{Discussion} \label{section 4.3}
The experimental results above support two facts: the OFI feature is more suitable than the original LOB, and the new Siamese network framework is more appropriate than the previous method, especially when utilizing the OFI features. In this section, we will analyze these two phenomena in more detail, examine the possible factors that affect the performance of baselines, and discuss the implications and limitations of our work.

\begin{figure}[htb]
    \centering
    \subfigure[\footnotesize The MAE performance plot (\textcolor{black}{\textbf{in red}}).]{
      \begin{minipage}[t]{0.47\textwidth}
        \includegraphics[width=\textwidth]{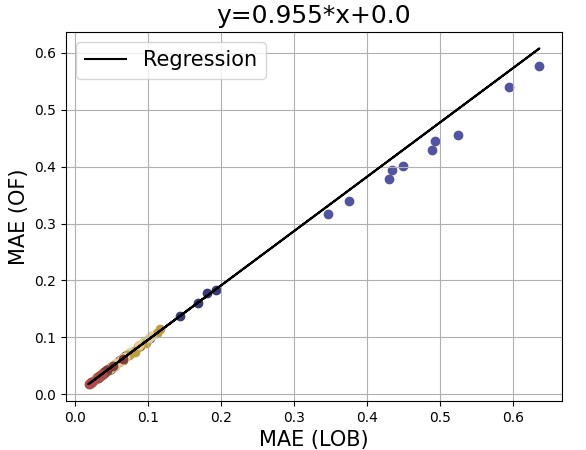}
        \label{tab2_a}
      \end{minipage}
      }
  \hfill
  \subfigure[\footnotesize The $R^2$ performance plot (\textcolor{cyan}{\textbf{in cyan}}).]{
      \begin{minipage}[t]{0.47\textwidth}
        \includegraphics[width=\textwidth]{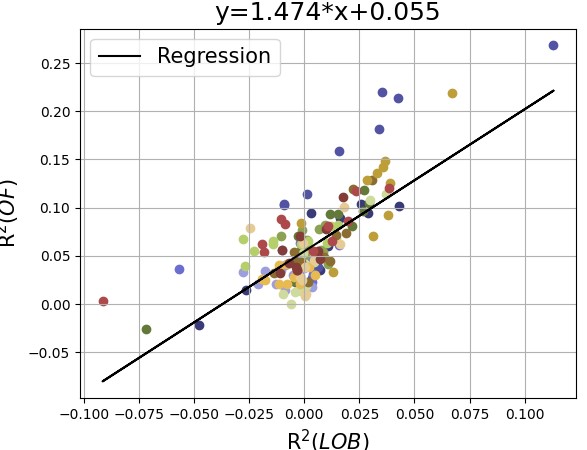}
        \label{tab2_b}
      \end{minipage}
      }
  \hfill
  \caption{\centering \small The performance comparison between LOB and OFI features.}
  \label{fig_tab2}
\end{figure}

Our first objective is to analyze the relationship between the model results based on Limit Order Book (LOB) and Order Flow Imbalance (OFI) features. We aim to understand why the OFI feature outperforms LOB. Taking inspiration from the research conducted by \citep{10}, we plan to calculate the average Mean Squared Error (MSE) and $R^2$ results of both LOB and OFI features on all test sets. Subsequently, we will perform a linear regression analysis, using these values as the explanatory and response variables, to further investigate the relationship between the two features.
The specific result is shown in Table \ref{table_2}, and we plotted some examples in Figure \ref{fig_tab2}.

It can be seen from the results that, for the MAE metrics, the intercept terms for all baselines are close to 0 except for MLP. And from the figures, we can see that, for MAE, the relationship between the two variables is more like a linear one. It may be because the network structure of training with LOB and OFI features is similar, and the training objective we adopted is to minimize the MAE metric. The slope of MAE is less than 1 and invariant, indicating the results utilizing the OFI features are better on various levels of data difficulty and are hardly affected by the horizon. For the $R^2$ metrics, the intercept terms for all baselines significantly exceed 0 except for MLP, the slope is larger than 1.0 at horizon 10, and it decreases as the horizon increases, demonstrating that the advantage of the OFI feature's interpretability to price changes is related to the predicted horizon. In the short horizon, the OFI feature has a greater advantage on hard data, while in the longer horizon, it has a greater advantage on simple data.

\begin{table}[htb]
  \tiny
  \caption{\centering The linear regression (slope/intercept) of utilizing the results of LOB and OFI features as explanatory and response variables, respectively.}
    \begin{tabular}{ccc|ccccc}
    \hline
    \multirow{2}{*}{Horizon} & \multirow{2}{*}{Metrics} & \multirow{2}{*}{Methods} & \multicolumn{5}{c}{Model Type} \\
          &   &  & MLP   & LSTM  & MLP-LSTM & CNN-LSTM & LSTM-MHA \\
    \hline
    \multirow{4}{*}{10} & \multirow{2}{*}{MAE} & Original & 0.945/0.005 & 0.972/0.0 & 0.963/0.0 & 0.964/0.001 & 0.969/0.0 \\
          &       & Siamese & 0.901/0.011 & 0.967/0.0 & 0.961/0.0 & 0.972/0.001 & \textcolor{black}{\textbf{0.955/0.0}} \\
          & \multirow{2}{*}{$R^2$} & Original & 0.353/-0.001 & 1.53/0.041 & 1.478/0.047 & 1.254/0.023 & 1.032/0.047 \\
          &       & Siamese & 0.4/-0.035 & 1.847/0.047 & 1.939/0.05 & 1.253/0.013 & \textcolor{cyan}{\textbf{1.474/0.055}} \\
    \hline
    \multirow{4}{*}{20} & \multirow{2}{*}{MAE} & Original & 0.94/0.01 & 0.962/0.0 & 0.948/0.001 & 0.959/0.001 & 0.967/0.0 \\
          &       & Siamese & 0.739/0.036 & 0.961/0.0 & 0.947/0.0 & 0.962/0.001 & 0.967/-0.001 \\
          & \multirow{2}{*}{$R^2$} & Original & 0.191/-0.099 & 1.04/0.058 & 1.103/0.068 & 0.936/0.048 & 0.873/0.057 \\
          &       & Siamese & 0.068/-0.39 & 1.465/0.067 & 1.534/0.082 & 1.22/0.029 & 0.915/0.077 \\
    \hline
    \multirow{4}{*}{50} & \multirow{2}{*}{MAE} & Original & 0.802/0.095 & 0.935/0.0 & 0.935/0.0 & 0.939/0.0 & 0.958/0.0 \\
          &       & Siamese & 0.939/0.024 & 0.948/0.0 & 0.942/0.0 & 0.942/0.0 & 0.97/-0.001 \\
          & \multirow{2}{*}{$R^2$} & Original & 0.467/-0.445 & 0.73/0.071 & 0.725/0.068 & 0.616/0.062 & 0.614/0.035 \\
          &       & Siamese & 0.803/-3.699 & 0.99/0.083 & 0.872/0.088 & 0.734/0.06 & 0.69/0.053 \\
    \hline
    \end{tabular}%
  \label{table_2}%
\end{table}%

\begin{figure}[htb]
    \centering
    \subfigure[\footnotesize The MAE performance plot (\textcolor{black}{\textbf{in red}}).]{
      \begin{minipage}[t]{0.47\textwidth}
        \includegraphics[width=\textwidth]{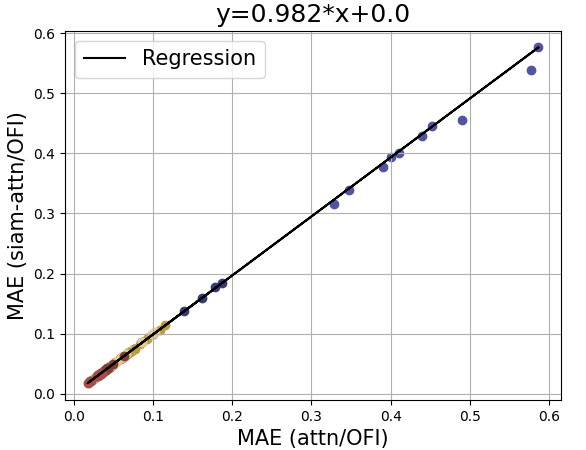}
        \label{tab5_a}
      \end{minipage}
      }
  \hfill
  \subfigure[\footnotesize The $R^2$ performance plot (\textcolor{cyan}{\textbf{in cyan}}).]{
      \begin{minipage}[t]{0.47\textwidth}
        \includegraphics[width=\textwidth]{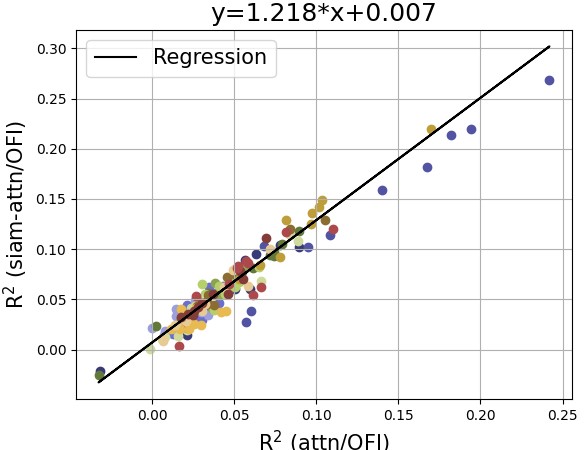}
        \label{tab5_b}
      \end{minipage}
      }
    \hfill
    \caption{\centering \small The performance comparison between Original and Siamese network.}
    \label{fig_tab3}
\end{figure}

Then, we want to analyze the relationship between the model results based on the original and novel Siamese network framework. Using the same method as above, we planned to calculate the average MSE and $R^2$ results of the original and Siamese network framework on all test sets and then carry out a linear regression analysis with these values as the explanatory and response variables. The specific result is shown in Table \ref{table_3}, and we plot some examples in Figure \ref{fig_tab3}.

The experimental results proved that for the MSE metrics, the relationship between outcomes corresponding to the LOB and OFI features is similar to that between the original and Siamese network methods. For the $R^2$ metrics, there is a more obvious linear relationship between them, and the performance improvement is more obvious when utilizing the OFI feature.

\begin{table}[htb]
  \tiny
  \caption{\centering The linear regression (slope/intercept) of utilizing the results of original and Siamese network framework as explanatory and response variables, respectively.}
    \begin{tabular}{ccc|ccccc}
    \hline
    \multirow{2}{*}{HistLen} & \multirow{2}{*}{Metrics} & \multirow{2}{*}{Feature} & \multicolumn{5}{c}{Model Type} \\
          &       &       & MLP   & LSTM  & MLP-LSTM & CNN-LSTM & LSTM-MHA \\
    \hline
    \multirow{4}{*}{10} & \multirow{2}{*}{MAE} & LOB   & 0.991/0.0 & 0.994/0.0 & 0.989/0.0 & 0.984/0.0 & 0.993/0.0 \\
          &       & OFI   & 0.726/0.043 & 0.991/0.0 & 0.983/0.0 & 0.985/0.0 & \textcolor{black}{\textbf{0.982/0.0}} \\
          & \multirow{2}{*}{$R^2$} & LOB   & 0.485/0.015 & 1.027/0.006 & 0.903/0.005 & 1.078/0.018 & 0.833/0.006 \\
          &       & OFI   & 0.38/-0.289 & 1.212/0.006 & 1.139/0.006 & 1.176/0.007 & \textcolor{cyan}{\textbf{1.218/0.007}} \\
    \hline
    \multirow{4}{*}{20} & \multirow{2}{*}{MAE} & LOB   & 1.006/-0.002 & 0.986/0.0 & 0.991/0.0 & 0.983/0.0 & 0.999/0.0 \\
          &       & OFI   & 0.551/0.065 & 0.983/0.0 & 0.982/0.0 & 0.994/0.0 & 0.991/0.0 \\
          & \multirow{2}{*}{$R^2$} & LOB   & 0.492/0.01 & 0.975/0.008 & 0.811/0.005 & 0.898/0.024 & 1.004/-0.001 \\
          &       & OFI   & 0.365/-0.456 & 1.212/0.008 & 1.117/0.013 & 1.136/0.006 & 1.12/0.012 \\
    \hline
    \multirow{4}{*}{50} & \multirow{2}{*}{MAE} & LOB   & 0.946/-0.001 & 0.984/0.0 & 0.991/0.0 & 0.992/-0.001 & 0.993/0.0 \\
          &       & OFI   & 0.533/0.105 & 0.987/0.0 & 0.992/0.0 & 0.985/0.0 & 0.987/0.0 \\
          & \multirow{2}{*}{$R^2$} & LOB   & 0.473/-0.01 & 0.875/0.005 & 0.741/-0.007 & 0.763/0.014 & 0.875/-0.012 \\
          &       & OFI   & 0.498/-2.351 & 1.058/0.01 & 0.964/0.019 & 0.971/0.01 & 0.976/0.014 \\
    \hline
    \end{tabular}%
  \label{table_3}%
\end{table}%

\begin{figure}[htb]
    \centering
    \includegraphics[width=1.0\linewidth]{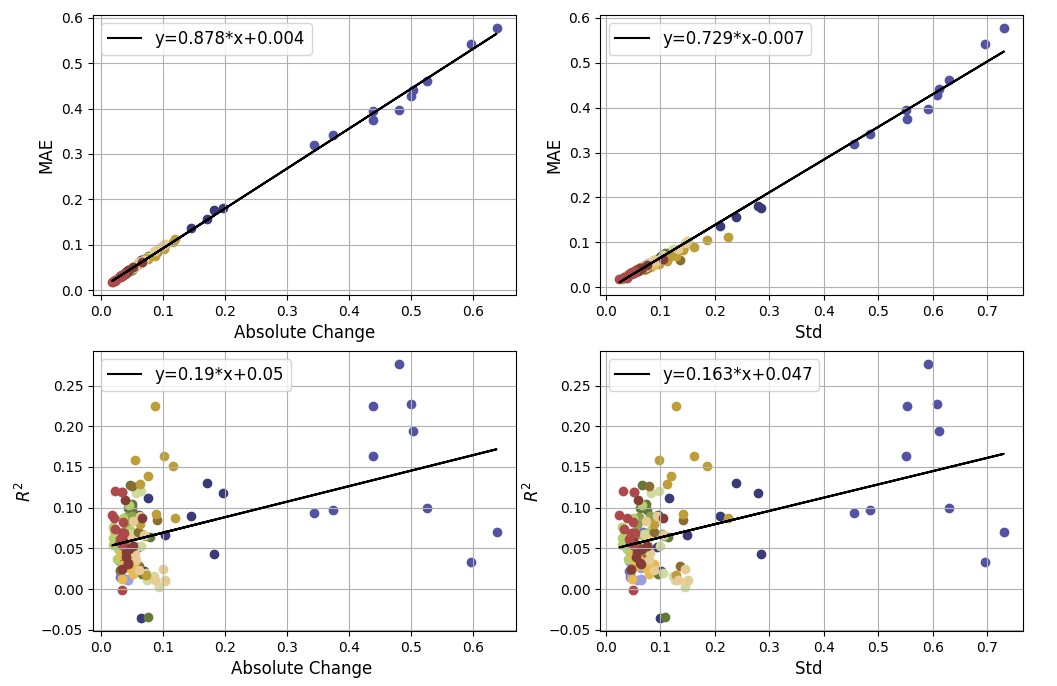}
    \caption{\centering \small The examples of performance plot (\textcolor{black}{in red}).}
    \label{fig_dist}
\end{figure}

Finally, we began with the data distribution attributes of the test set to explore the factors that affect model performance. We believed that stock price volatility on the test set will affect the prediction results, so we wanted to study the correlation between stock price volatility and model performance. To estimate if the data volatility would influence the performance of baselines, we proposed two indicators on the test set: average absolute change (AC) and standard deviation (std) and \textcolor{black}{analyzed} whether there is a correlation between the two and the model performance. The specific result is shown in Table \ref{table_4}, and we plotted some examples in Figure \ref{fig_dist}.

\begin{align}
    AC_{i} = &\frac{1}{N_i}\sum_{j=1}^{N_i}\lvert p_{ij}-\frac{1}{N_i}\sum_{j=1}^{N_i}p_{ij}\rvert  \\
    std_{i} = &\sqrt{\frac{\sum_{j=1}^{N_i}(p_{ij}-\frac{1}{N_i}\sum_{j=1}^{N_i}p_{ij})^2}{N_i}}
\end{align}
where $N_i$ is the data number of test set $i$, $p_{ij}$ is the label of $j^{th}$ data. It can be seen from the results that MAE has a strong positive linear relationship with data volatility, and $R^2$ also has a weak positive correlation with data volatility. It can be considered that the explainable part only accounts for a small proportion of the actual price changes, and the remaining unexplainable component is the effect of high noise, which is the source of the difficulty of this prediction task.

\begin{table}[h]
  \tiny
   \caption{\centering The linear regression (slope/intercept) of the data statistics and model performance as explanatory and response variables, respectively.}
    \begin{tabular}{ccc|cccc}
    \hline
    \multirow{2}{*}{HistLen} & \multirow{2}{*}{Feature} & \multirow{2}{*}{Architecture} & \multicolumn{4}{c}{Regression Type} \\
          &       &       & AC-MAE & Std-MAE & AC-$R^2$ & Std-$R^2$ \\
    \hline
    \multirow{4}{*}{10} & \multirow{2}{*}{LOB} & Original & 0.993/0.001 & 0.823/-0.012 & 0.041/-0.006 & 0.039/-0.007 \\
          &       & Siamese & 0.983/0.001 & 0.816/-0.012 & 0.06/-0.001 & 0.054/-0.003 \\
          & \multirow{2}{*}{OFI} & Original & 0.91/0.003 & 0.756/-0.009 & 0.188/0.033 & 0.158/0.03 \\
          &       & Siamese & \textcolor{black}{\textbf{0.878/0.004}} & \textcolor{black}{\textbf{0.729/-0.007}} & \textcolor{black}{\textbf{0.19/0.05}} & \textcolor{black}{\textbf{0.163/0.047}} \\
    \hline
    \multirow{4}{*}{20} & \multirow{2}{*}{LOB} & Original & 0.996/0.002 & 0.826/-0.011 & 0.101/-0.037 & 0.094/-0.039 \\
          &       & Siamese & 0.99/0.002 & 0.821/-0.011 & 0.128/-0.041 & 0.119/-0.044 \\
          & \multirow{2}{*}{OFI} & Original & 0.9/0.005 & 0.748/-0.007 & 0.223/0.017 & 0.188/0.014 \\
          &       & Siamese & 0.883/0.005 & 0.734/-0.007 & 0.163/0.039 & 0.141/0.036 \\
    \hline
    \multirow{4}{*}{50} & \multirow{2}{*}{LOB} & Original & 1.01/0.006 & 0.838/-0.008 & 0.405/-0.192 & 0.377/-0.202 \\
          &       & Siamese & 1.008/0.005 & 0.836/-0.008 & 0.41/-0.187 & 0.38/-0.197 \\
          & \multirow{2}{*}{OFI} & Original & 0.921/0.007 & 0.766/-0.005 & 0.349/-0.088 & 0.311/-0.095 \\
          &       & Siamese & 0.911/0.008 & 0.758/-0.005 & 0.288/-0.07 & 0.257/-0.076 \\
    \hline
    \end{tabular}%
  \label{table_4}%
\end{table}%

Unlike previous work, this paper focuses on the deep learning model in the Chinese A-share market. We employed the deep learning model, adopted the original LOB data and derived OFI features, and added the MHA mechanism to experiment on 14 defense-industry-related stocks. Some conclusions are consistent with those in the Nasdaq share market, but there were some differences. For example, CNN-related networks had no advantage in feature extraction in the A-share market. The further processing of features by MHA improves performance. In addition, the most vital contribution of this paper is to propose a new Siamese network framework based on the symmetry of LOB data itself. This framework has good universality and adapts well to various network architectures. Experimental results show that our proposed architecture has significantly improved diverse models and two kinds of input features. This also inspires related work to construct a more appropriate deep-learning model for this prediction task based on the traits of the data itself and apply our ideas to further improve the results. \textcolor{black}{The Siamese network framework leverages the symmetry traits of LOB data, which allows the model to better capture the inherent structure and relationships within the data, thereby enhancing its predictive capabilities. The MHA mechanism, when applied to the OFI features, further enhances the model's ability to capture complex patterns and dependencies, leading to improved performance.}

\textcolor{black}{Although we have reached some useful conclusions, the study in this paper has some flaws. First, due to limitations in research resources, the validity of this structure in the U.S. and European stock markets has not been verified; second, the features used in this study are relatively simple, and we have not validated the model's capabilities on more complex features; last but not least, we have not further studied how to exploit the results to develop practical profit-making strategies in the A-share market, which is beyond the scope of this paper and will be taken as our next goal.}

\section{Conclusion}
\textcolor{black}{
Our research has been dedicated to addressing the challenges of price forecasting using Limit Order Book (LOB) data, which is characterized by its high-dimensional and dynamic nature. The primary motivation behind this study was to leverage the symmetry traits of LOB data to enhance forecasting accuracy. To achieve this, we proposed a novel approach utilizing the Siamese architecture, which processes the ask and bid sides of the LOB data using the same module with parameter sharing. This method not only maintains the symmetry of the data but also improves data efficiency. We tested the Siamese-based methods on several strong baseline models with different forecasting horizons and demonstrated their effectiveness on the Chinese A-share market. Our results showed that the proposed architecture significantly improved the performance of various models, regardless of whether the input was the original LOB data or the Order Flow Imbalance (OFI) features. Additionally, we combined the Multi-Head Attention (MHA) mechanism with the LSTM module and found that MHA could further enhance model performance, particularly over short forecasting horizons.}

\textcolor{black}{The significance of our study lies in its ability to effectively utilize the unique characteristics of LOB data, providing a fresh perspective for enhancing model performance. By leveraging the symmetry traits of LOB data, our method demonstrates strong generalization capabilities, which can be applied to various network architectures. This approach not only improves forecasting accuracy but also offers a new avenue for research in high-frequency trading (HFT) and financial market analysis.}

\textcolor{black}{Looking ahead, there are several promising directions for future work. First, we plan to explore the role of symmetry traits in more complex features. This will help us better understand how to further enhance model performance. Second, we aim to investigate the differences between the A-share market and the U.S. and European stock markets. Given the unique characteristics of each market, it is essential to determine whether our proposed methods can be generalized or if they need to be adapted to suit different market conditions. Finally, we intend to focus on developing practical profit-making strategies based on our experimental results. This will involve determining the optimal forecasting horizon and designing suitable trading strategies that can effectively utilize the predicted price movements to generate profits in the HFT process.}

\textcolor{black}{In conclusion, our study has made significant contributions to the field of financial market analysis by proposing a novel Siamese network framework that effectively leverages the symmetry traits of LOB data. The results of our research not only enhance the performance of existing deep learning models but also provide valuable insights for future work in this area. We are excited about the potential of our methods to be applied in more complex models and different market contexts, and we look forward to further exploring their practical applications in HFT and other financial domains.}

\clearpage

\section{Abbreviations}

\noindent High-Frequency Trading \qquad       HFT

\noindent Limit Order Book\qquad              LOB

\noindent Multi-Head Attention\qquad          MHA

\noindent Order Flow Imbalance\qquad          OFI

\noindent Support Vector Machine\qquad        SVM

\noindent Least Mean Square\qquad             LMS

\noindent Multi-Layer Perceptron\qquad        MLP

\noindent Convolutional Neural Network\qquad  CNN

\noindent Long Short-Term Memory\qquad        LSTM

\noindent Gated Recurrent Network\qquad       GRU

\noindent Ordinary Least Square\qquad         OLS

\noindent Mean Absolute Error\qquad           MAE

\noindent Mean Square Error\qquad             MSE

\section{Acknowledgement} 
This work has been supported by The Youth Innovation Promotion Association of the Chinese Academy of Sciences (E1291902), Jun Zhou (2021025).

\section{Data Availability Statements}
The datasets generated during and/or analyzed during the current study are available from the corresponding author upon reasonable request.

\section{Declarations}
\textbf{Informed consent} \qquad Informed consent was obtained from all individual
participants included in the study.

\textbf{Conflict of Interests} \qquad The authors declare that they have no known competing financial interests or personal relationships that could have appeared to influence the work reported in this paper.


\clearpage

\bibliography{sample.bib}

\clearpage
\appendix
\section{Details of Baselines} \label{baselines}
This paper adopts a variety of baseline models based on deep learning. The detailed information on the models is summarized as follows.

\begin{itemize}
    \item The MLP model has 3 hidden layers, and the numbers of hidden neurons are 500, 250, and 64 respectively. The input dim is $40\times 50 = 2000$ for the original LOB as input, and 1000 for the OFI features.
    \item The stacked LSTM network contains 3 layers of LSTM, and all the hidden size is equal to 64. Then, all the final outputs of the three layers are concatenated and fed to a linear layer, so the inputs are $64\times 3 = 192$. The input dim is 40 for the original LOB as input and 20 for the OFI features.
    \item Compared to the stacked LSTM above, MLP-LSTM adds a single hidden layer MLP module before the input of the stacked LSTM to process the input data. The hidden neuron number of the MLP is 128, and the output dim is 64.
    \item Compared to the MLP-LSTM above, CNN-LSTM replaces the MLP module with an inception module. The inception module consists of 1-dimensional convolution kernels with sizes of 1, 3, 5, and 7, and the output size is 16. Then, all the outputs are concatenated as 64-dimensional inputs to be fed into the LSTM module.
    \item Compared to the stacked LSTM above, the LSTM-MHA replaces the final layer of LSTM with a multi-head attention mechanism, so it contains two LSTM layers and an attention layer. The final outputs of LSTM and attention are concatenated as the input of a linear projection layer.
    \item We also separately tried the stacked Multi-Head Attention layers with causal (the decoder part of the Transformer) and non-causal (the encoder part of the Transformer) architecture of 4 layers, and the hidden dim is equal to 64, but both models performed poorly.
\end{itemize}

The network structure based on the Siamese network framework is consistent with the baseline comparison, except that the input dimension is halved.

\end{document}